\documentclass[economies,review,accept,pdftex,moreauthors]{Definitions/mdpi} 

\firstpage{1} 
\pubvolume{1}
\issuenum{1}
\articlenumber{0}
\pubyear{2024}
\copyrightyear{2024}
\datereceived{ } 
\daterevised{ } 
\dateaccepted{ } 
\datepublished{ } 
\hreflink{https://doi.org/} 

\usepackage{wrapfig}
\usepackage{graphicx}%
\usepackage{multirow}%

\Title{Resource Allocation with Karma Mechanisms}

\TitleCitation{Resource Allocation with Karma Mechanisms}

\Author{Kevin Riehl $^{1,\dagger}$*\orcidA{}, Anastasios Kouvelas $^{1}$\orcidB{} and Michail Makridis $^{1,2}$\orcidC{}}

\AuthorNames{Kevin Riehl, Anastasios Kouvelas and Michail Makridis}

\AuthorCitation{Riehl, K.; Kouvelas, A.; Makridis, M.}

\address{%
$^{1}$ \quad Traffic Engineering Group, Institute for Transport Planning and Systems, ETH Zurich, Stefano-Franscini-Platz 5, Zurich, 8093, Switzerland; kriehl@ethz.ch \& kouvelas@ethz.ch \\
$^{2}$ \quad Transport and Traffic Engineering Group, Institute of Data Analysis and Process Design, Zurich University of Applied Sciences, Technikumstrasse 81, Winterthur, 8400, Switzerland; makr@zhaw.ch}

\corres{Correspondence: kriehl@ethz.ch }

\abstract{
Monetary markets serve as established resource allocation mechanisms, typically achieving efficient solutions with limited information. 
However, they are susceptible to market failures, particularly under the presence of public goods, externalities, or inequality of economic power. 
Moreover, in many resource allocating contexts, money faces social, ethical, and legal constraints. 
Consequently, research increasingly explores artificial currencies and non-monetary markets, with Karma emerging as a notable concept. 
Karma, a non-tradeable, resource-inherent currency for prosumer resources, operates on the principles of contribution and consumption of specific resources. 
It embodies fairness, near incentive compatibility, Pareto-efficiency, robustness to population heterogeneity, and can incentivize a reduction in resource scarcity. 
The literature on Karma is scattered across disciplines, varies in scope, and lacks of conceptual clarity and coherence.
Thus, this study undertakes a comprehensive review of the Karma mechanism, systematically comparing its resource allocation applications and elucidating overlooked mechanism design elements. 
Through a systematic mapping study, this review situates Karma within its literature context, offers a structured design parameter framework, and develops a road-map for future research directions.
}

\keyword{Resource Allocation, Artificial Currency, Non-monetary economics, Systematic Mapping Study, Karma} 

\JEL{C71, D44, D61, H41}

\begin{document}


\section{Introduction}\label{sec1}

Resource allocation mechanisms are decision making processes about distributing limited and scarce resources among a group of recipients~\citep{paccagnan2022utility}.
Central resource allocation involves one deciding entity, which requires a vast amount of information for this task.
On the contrary, market mechanisms employ a distributed decision-making process, in which agents decide on their resource-allocating actions autonomously according to their subjective utility.
Markets prove to be powerful, as they achieve near Pareto-efficient solutions with a minimal amount of information~\citep{piccoli2023control}.
While barter markets are of limited practicality, monetary markets are a useful tool for resource allocation in practice, due to the use of a standardized, globally-accepted, monetary currency that can be traded in for any resource~\citep{kregel2021economic}. 

However, money does not always do its job, and monetary markets fail sometimes;
phenomena such as inequality of economic power, public goods, and externalities, to name a few, have been observed to lead to undesirable, inefficient allocations \citep{coase1960problem,hardin1968tragedy,bator1995anatomy,peterson2017economic}.
What's more, in many contexts, the use of money is not desired, socially not accepted, considered unethical, or even prohibited (e.g. donations, shared resources, collective decision-making within organizations)~\citep{article_organ_A}.
A growing branch of literature is concerned with artificial currencies~\citep{gorokh2021monetary}, which represent non-monetary markets and resource allocation mechanisms.
For isolated, single-stage resource allocation problems, extensive work on non-monetary matching and combinatorial assignment problems has been conducted \citep{budish2011combinatorial,dutot2011approximation,skowron2013non,carreras2016theory,nguyen2016assignment,karaenke2020non}.
For repeated resource allocation problems, there are few works on non-monetary market mechanisms yet, among which Karma has evolved as an important narrative.

Karma employs a currency different from money; it can only be gained by producing and only be lost by consuming a specific resource.
It is a resource-inherent, non-monetary, non-tradeable, artificial currency for prosumer resources (produced and consumed by market participants likewise).
As a non-monetary mechanism, Karma complements monetary markets and provides attractive properties. 
For example, it is fairness-enhancing, near incentive-compatible, and robust towards population heterogeneity \citep{article_567,article_1002,article_600}. 
Due to the design of Karma, one can consider Karma as playing against your future self, as the only way to consume is to put effort and produce first, and future needs must be traded off against present needs when consuming.
Last but not least, Karma is reported to not only concern the efficiency and fairness of resource allocation but also to lead to a decrease in resource scarcity in peer-to-peer markets \citep{article_foodbank,article_organ_A,article_organ_B,article_organ_C}.

Let us discuss Karma at the example of public goods, which are shared resources, that have unrestricted access to consumption, and a diminishing marginal utility with increasing consumption for the users. 
The road network of a city is a suitable example for shared, public goods. 
Road users are free to drive into the city, and this is useful as users can reach work, shopping, social, living, and recreational facilities faster when compared to other modalities. 
However, if too many drive by, congestion arises, and nobody moves a single inch. 
A well-discussed countermeasure is the introduction of monetary markets, namely: congestion pricing~\citep{de2011traffic}, where vehicles need to pay a surcharge for driving in the city. 
This leads to a decrease in the demand for the network, and to sustainable, congestion-free levels of usage. 
Due to the monetary disincentive, people will trade off their urgency to drive to the city, and their willingness to pay.
However, congestion pricing can be problematic, as equity issues emerge in a society with unequal distribution of economic power: the poorest will most likely not be able to afford the charge, and thus consume significantly less.
During implementation, equity considerations are the major hurdle for congestion pricing to overcome in the political debate.
Karma could make a difference here: not driving could be considered as producing, and driving could be considered as consuming the resource "right of driving to the city". 
Instead of paying money as a congestion pricing tax, Karma points could be used. 
These Karma points cannot be bought, but only gained by not consuming.
This would force individuals not to trade off the price with other resources they could buy alternatively but to solely consider present versus future consumption of this specific resource.
What's more, the socio-economic contexts, such as income or wealth, and therefore the above-mentioned equity considerations, would not play a role anymore.

The literature on Karma is strongly dispersed over different fields and years. 
Starting with the first publication in 2003, Karma applications targeted issues on peer-to-peer computer network markets for filesharing, where free-riding (consuming files without providing content for others) was a major issue.
From 2014, Karma became an important technological component for modern blockchain technologies such as crypto currencies, and a fairness-enhancing incentive mechanism in behavioral studies.
Beginning from 2019, Karma gained momentum as an artificial currency and resource allocation mechanism in the economic discourse, and is modelled as a dynamic population game.
Consequently, discussions on Karma vary in scope, lack methodological consistency and conceptual clarity, as particular mechanism design aspects might be overlooked.
The diversity of Karma applications highlights the need for a systematic framework of design parameters \& options.
A comprehensive overview of Karma could enhance more systematic research.

Therefore, this work aims to provide an exhaustive literature review of the Karma mechanism.
The review covers an exhaustive overview and systematic comparison of Karma applications as resource allocation mechanism, a Karma framework with design parameters and options, discusses its game theoretic modelling, and provides future research directions.
Altogether, the contributions of this review can be summarized as follows:
This review connects Karma with the context of its literature, which enables the perception of Karma as a non-monetary, resource allocation mechanism to a broader audience.
This review provides a unifying framework of existing Karma mechanism designs, which enables more systematic analysis in future studies.

The remainder of this work is as follows.
Section~\ref{structure} elaborates on the origins of the Karma mechanism and depicts the intellectual structure \& trends of the Karma citing literature.
Section~\ref{mechanism} reviews the applications of Karma as a resource allocation mechanism and distills relevant mechanism design elements.
Section~\ref{future} elaborates on promising future research directions for Karma related works.
Finally, Section~\ref{methods} presents the research protocol and outlines the methodologies used for the literature review (systematic mapping study, Latent Dirichlet Allocation).
The study closes with concluding remarks on Karma in Section~\ref{conclusion}.

\section{Review of the Karma Literature}\label{structure}


\subsection{The origins of Karma}
The original work \textit{Karma: A secure economic framework for peer-to-peer resource sharing} \citep{vishnumurthy2003karma} provides a secure algorithm and protocol for resource sharing in peer-to-peer networks. 
The authors were inspired by the religious concept of Karma in Hinduism, where morally good actions will be rewarded and morally bad actions will be punished~\citep{larson2020karma}.
One of the major motivations for the original work was to solve the free-riding problem on peer-to-peer network marketplaces, which is the majority of users consume but only a few provide resources. 
The authors demonstrate, that their proposed concept forces participants to achieve a parity between resource contribution \& consumption, and prove Karma's protocol properties such as non-repudiation, certification, and atomicity.

Even though Karma was originally intended for resource sharing in peer-to-peer networks only, 
it was used as a network protocol for secure, attack-proof information dissemination systems across peer-to-peer networks.
Also, it is considered the first decentralized crypto currency, based on a proof-of-work minting\endnote{\url{https://medium.com/blockwhat/03-it-s-karma-484fdc2d8657}.}, which is a foundational technology for modern blockchain technologies.
Besides, Karma was discussed as an accounting system, 
and its behavioral implications were discussed as an incentive mechanism.
In addition to that, Karma was modeled as a dynamic population game, 
and discussed as a resource allocation mechanism
~\citep{caldarola2022neural,antoniadis2005incentives,article_600,zhao2009analysis}.

\begin{figure}[H]
    \begin{adjustwidth}{-\extralength}{0cm}
    \centering
    \includegraphics[width=15.5 cm]{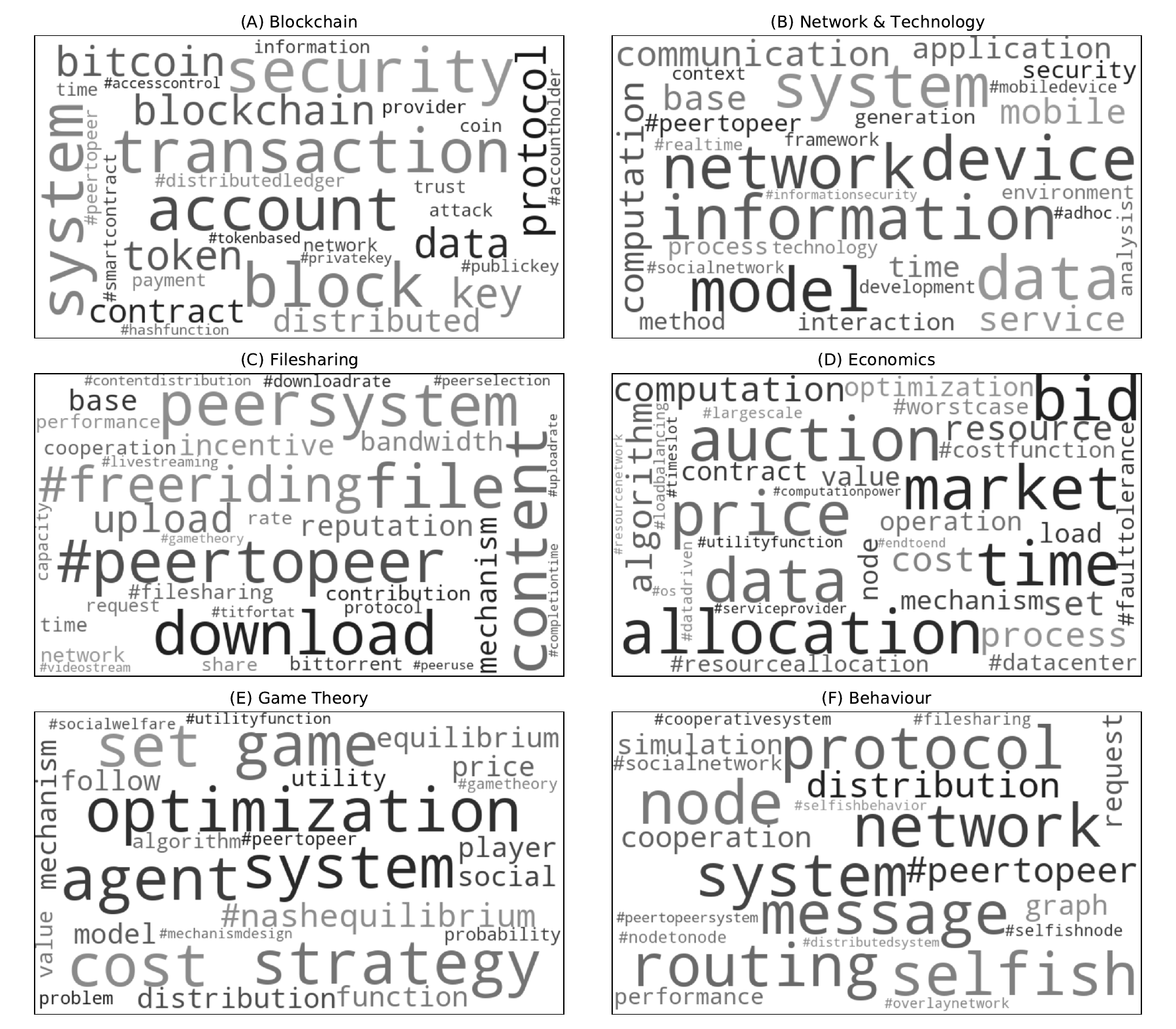}
    \end{adjustwidth}
    \caption{
        \textbf{Clusters in the literature corpus.}  
        This figure shows word clouds of selected, most frequent words for the six identified topic clusters. The size of words relates to their frequency in the documents. 
        }
    \label{fig-wordclouds}
\end{figure}

\begin{figure}[H]
    \begin{adjustwidth}{-\extralength}{0cm}
    \centering
    \includegraphics[width=15.5 cm]{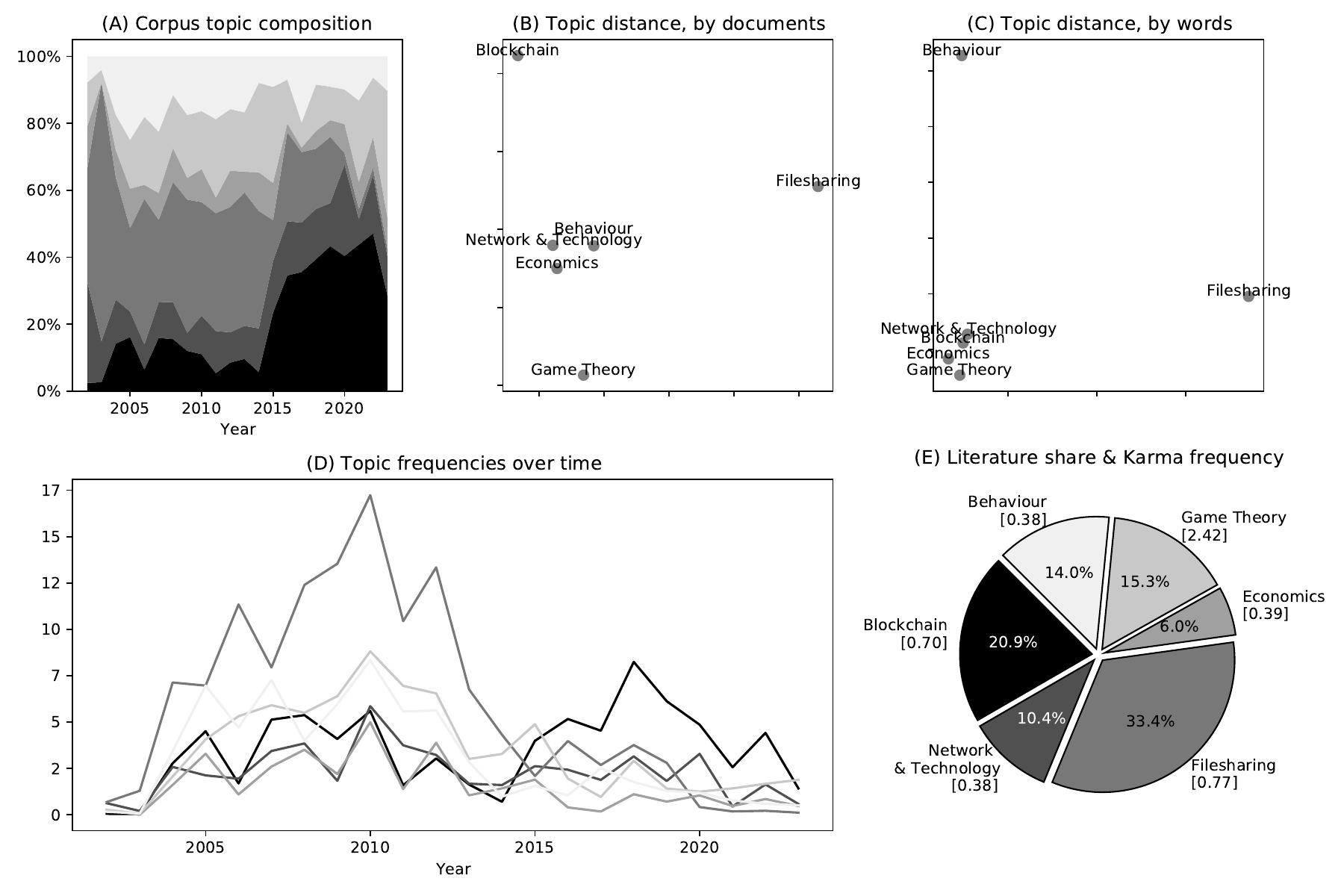}
    \end{adjustwidth}
    \caption{
        \textbf{Composition \& structure of literature corpus.} 
        (A) This figure shows corpus' topic composition. 
        Share of citing literature for a specific year is determined by LDA document-topic matrix probabilities.
        (B+C) These charts express similarity of topics depicted as topic distance by documents resp. by words (PCA of LDA's matrices into 2D).
        (D) This figure shows number of primary citations of Karma for each topic and year.  
        (E) The pie chart displays share each topic has in the literature on average over observed period of two decades. 
        Number in brackets describes how often "Karma" appears on average in works of this topic.
        } 
    \label{fig-corpus}
\end{figure}

\begin{figure}[H]
    \begin{adjustwidth}{-\extralength}{0cm}
    \centering
    \includegraphics[width=15.5 cm]{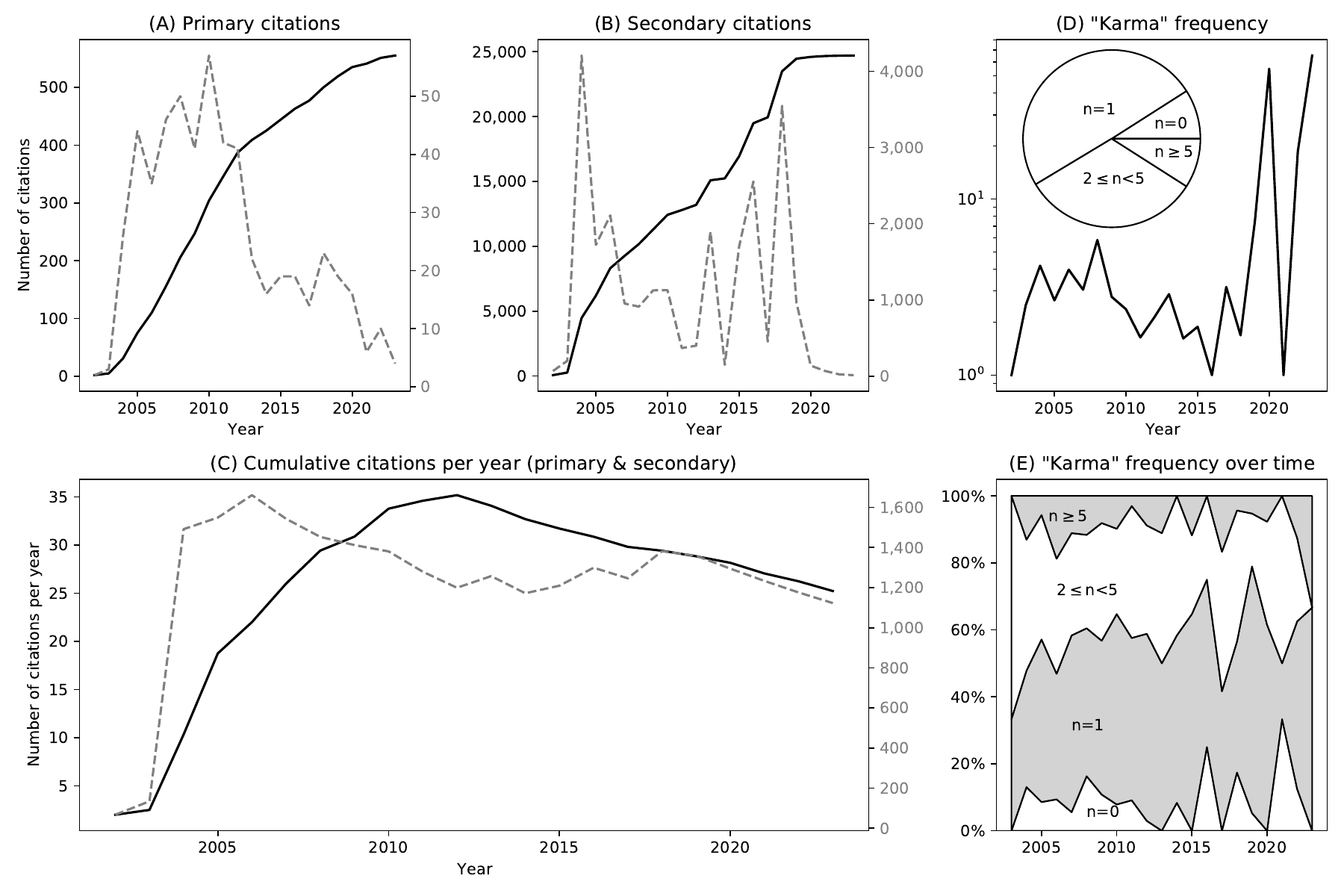}
    \end{adjustwidth}
    \caption{
        \textbf{Two decades of Karma.} 
        (A) This figure shows primary citations (dashed line = citations per year with ordinate on right, black line = cumulative citations with ordinate on left). 
        (B) This figure shows secondary citation.
        (C) This diagram shows number of cumulative primary (black) and secondary (dashed) citations divided by the number of years after publication.
        (D+E) These charts show number of times "Karma" appears per document, excluding the references.
        It indicates Karma's role in a paper (only citation or more intensively worked with concept).
        } 
    \label{fig-overtime}
\end{figure}

\subsection{Two decades of Karma}
The analysis of the Karma citing literature reveals, that there are six distinct clusters within the Karma citing literature: "Blockchain", "Network \& Technology", "Filesharing", "Economics", "Game Theory", and "Behaviour" (Fig.~\ref{fig-wordclouds}).
Figure~\ref{fig-corpus}(A) explains, that "Filesharing" dominated the Karma literature at the beginning, and that beginning from 2015 "Blockchain" started to subjugate.
The analysis of topic distance in Figure~\ref{fig-corpus}(B+C) outlines,
that "Blockchain", "Filesharing" and "Game Theory" stand out (document-distance),
and that "Behaviour" and "Filesharing" (word-distance) differ from the other topics.
With regards to the literature share of each topic in Figure~\ref{fig-corpus}(E), one can see that the largest share ($\sim$30\%) of the literature is on "Filesharing", and the second largest share ($\sim$20\%) is on "Blockchain".

Over the past two decades, Karma gained momentum as a concept since its first appearance, until its climax 2012.
Figure~\ref{fig-overtime} shows how Karma evolved in the literature over the past two decades.
With regards to the primary citations in Figure~\ref{fig-overtime}(A), we can observe that around 400 works per year cited the original Karma paper from 2005 to 2010.
Afterwards, the number declined to around 200 works per year. From 2020, the number of primary citations dropped. 
To date, Karma was cited around 583 times. 
With regards to the secondary citations in Figure~\ref{fig-overtime}(B), we can observe that the number of secondary citations increased strongly around 2005 and then later between 2015 and 2020, with around 15,000 secondary citations per year. 
To date, Karma was secondary cited around 25,000 times. 
With regards to the cumulative citations divided by years since publication in Figure~\ref{fig-overtime}(C), we can see that the number of average primary citations since publication is steadily increasing until 2012 to 35 citations per year, and then slightly decreases since then. 
A similar observation can be made for the average secondary citations since publication. 

An increasing amount of works started to work more deeply with the concept of Karma beginning from 2017.
With regards to the number of times the term "Karma" is used in the papers in Figure~\ref{fig-overtime}(D), we can better understand what drives the number of primary and secondary citations. 
To start with, we can observe that "Karma" was used as a term in scientific studies around 2-3 times per document on average until 2017. 
Afterwards instead, it strongly increased up to 14 times (with an exception in 2021). 
This development indicates, that while "Karma" was just cited as a side note at the beginning, it was recognized and developed further as a concept later on. 
Therefore, one could argue that interest in the concept of "Karma" increased since 2017, as the citing literature started to work with the concept more intensively. 
The pie chart makes it obvious that the peaks in Figure~\ref{fig-overtime}(D) are mainly driven by few publications, as the vast majority (60\%) of the documents cited Karma only 1-2 times.
With regards to Figure~\ref{fig-overtime}(E) this can be supported by an increase in the share of publications up to 30\% that cites "Karma" more than 4 times on average per document.

First, Karma was prominent in "Filesharing" field, then the "Blockchain" field, and now the "Game Theory" field predominantly works with Karma.
Figure~\ref{fig-corpus}(A) exhibits the fields that drive the citations of the original Karma paper. One can see that while at the beginning "Filesharing" was predominantly citing (around 40\%) in the first years (2002-2014), there was a shift towards "Blockchain" (around 35\%) related parts of the citing literature corpus (beginning from 2014); in the last years (2022-2023) one can see a significant increase in "Game Theory" citations (around 25\%). 
With regards to the absolute number of citations in Figure~\ref{fig-corpus}(D) one can see that the "Filesharing" literature was around two times larger than the "Blockchain" literature, and the emerging "Game Theory" literature is even smaller.
Figure~\ref{fig-corpus}(E, number in brackets) displays how frequently the term Karma was used per document in the different fields. One can see that the Karma frequency as discussed in Figure~\ref{fig-overtime}(D) was mainly driven by the "Game Theory" field, where on average each paper cited the term "Karma" 2.42 times, compared to "Filesharing" (0.77 times) and "Blockchain" (0.70 times).

\subsection{Karma modelled as a game}


Karma is described inconsistently in its applications as a resource allocation mechanism.
It is described verbally, rather than modelled formally for quantitative analysis.
When simulating Karma (e.g. as a multi-agent system), assumptions on the agents' behaviour is made at many times.
In 2019, Karma was modelled with a game theoretical formalism~\citep{article_600}, which enabled a more precise, reproducible, quantitative assessment of this novel resource allocation mechanism.
The benefit of this formalism is the possibility to predict the behaviour of selfish (rational) agents, when simulating Karma resource allocation systems.

Karma is described as a repeated, stochastic, dynamic population game. 
It applies to games that are not played once, but multiple times (repeated).
Many aspects of the game are modelled probabilistically (stochastic), i.e. the bidding behavior is modelled as Markov decision process.
The formalism applies to dynamic population games, in order to represent the strategic interplay in large societies of rational (selfish) agents~\citep{article_1001}.

In the Karma game, there is a population of agents, where each agent...
\begin{itemize}
    \item has a specific amount of Karma
    \item has a random, time-varying urgency (represents the agent's cost when not getting a specific resource)
    \item has an individual temporal consumption preference type (discount factor, represents the subjective trade-off between consuming now versus later)
\end{itemize}
Agents are matched randomly every round to compete for a specific resource, by bidding with Karma in auction-like setups.
Depending on their urgency, Karma balance, and consumption type, agents must determine an optimal bid to earn the resource when necessary, while accounting for potential future competitions in following rounds.
This bidding behavior can be described by a probabilistic distribution called the policy.
The optimal policy describes the behavior of rational agents, meaning that no agent can gain by deviating from this policy.
Moreover, the population's Karma balance distribution (across urgency, Karma balance, and consumption types) can be described by a probabilistic distribution called the state.
Over time, a population of rational agents will converge to a stationary state called the stationary Nash equilibrium.
The stationary Nash equilibrium consists of the optimal policy and stationary state distribution, is guaranteed to exist for dynamic population games, and can be calculated. 
The interested reader is highly recommended to review further details in the works of ~\cite{article_1001,article_1002,article_600}.

\section{Karma Resource Allocation \& Mechanism Design}\label{mechanism}

A growing number of works study and employ Karma as a resource allocation mechanism for prosumer resources.
Often, Karma is used for the distribution of resources in peer-to-peer networks.

An exhaustive compilation of relevant Karma applications for resource allocation mechanism can be found in Tables~\ref{table_applications_A}+\ref{table_applications_B}+\ref{table_applications_C}.
In the literature field "Filesharing", Karma was applied in virtual, P2P networks, for the allocation of computational resources such as files~\citep{vishnumurthy2003karma,vishnumurthy2008substrate,article_411} and computation power~\citep{article_243,article_492}. 
In the literature field "Network \& Technology", Karma was applied in telecommunication networks, for mitigating distortion signal interference with other agents~\citep{article_185}, and sharing of internet access~\citep{article_273,article_348,article_55,concept_nuglets1,concept_nuglets2}.
In the literature field "Blockchain", Karma was not applied as mechanism, as this branch of literature mainly focuses on applications as accounting system for blockchain technologies such as crypto currencies.
In the literature field "Game Theory", Karma was applied in road transportation networks, for toll-pricing~\citep{article_265,article_z02}, intersection management~\citep{article_567,article_600}, and priority lanes~\citep{article_1002,article_1004}.
In the literature fields "Behaviour" and "Economics", Karma was applied in social and logistic networks, for the distribution of food donations~\citep{article_foodbank}, babysitting services~\citep{article_scrips}, living-organ donations~\citep{article_organ_A,article_organ_B,article_organ_C}, and vacancies for individuals~\citep{article_college_A,article_college_B}.

\begin{table}[H]
    \caption{\textbf{Karma mechanism design parameters \& options}}
    \label{table_design_parameters}
    \begin{tabularx}{\textwidth}{ll}
        \toprule
        \textbf{Design Parameter} & \textbf{Option} \\
        \midrule
        \textbf{Currency}  & \\
            \hspace*{0.5cm} Parity         & Price, Threshold, Binary \\
            \hspace*{0.5cm} Balance limits & Unlimited, Bounded (upper, lower) \\
            \hspace*{0.5cm} Amount control & Constant, Constant per capita, Uncontrolled, Expiry \\
            \hspace*{0.5cm} Initialization & Equal endowment, Weighted endowment, Random endowment, None \\
            \hspace*{0.5cm} Redistribution & Property tax, Payment tax, Lottery, Expiry, None \\
         & \\
        \midrule
        \textbf{Interaction}  & \\
            \hspace*{0.5cm} Price control & Auction, Centrally defined, None \\
            \hspace*{0.5cm} Price limits  & Only positive, Fix , None \\
            \hspace*{0.5cm} Resource provision & By one agent, By all agents, By system \\
            \hspace*{0.5cm} Resource allocation & Auction winner, System decision, Provider decision \\
            \hspace*{0.5cm} Counter-party & N agents, One agent, System \\
            \hspace*{0.5cm} Peer selection & Market, Neighbourhood, Randomly assigned, Active selection \\
            \hspace*{0.5cm} Decision-making & Free \\
            \hspace*{0.5cm} Urgency process & Homogeneous, Heterogeneous \\
            \hspace*{0.5cm} Temporal preference & Homogeneous, Heterogeneous \\
         & \\
        \midrule
        \textbf{Transaction}  & \\
            \hspace*{0.5cm} Payment amount & Bid, Peer's bid, Difference in bids, Fixed, Nothing \\
            \hspace*{0.5cm} Payment receiver & \parbox[t][][t]{9cm}{Resource provider, System, Equally across population, Weighted across population} \\
            \hspace*{0.5cm} Karma gain & Resource provision, Resource consumption  \\
            \hspace*{0.5cm} Karma loose & Resource consumption, Expiration, Rule-violation  \\
         & \\
        \bottomrule
		\end{tabularx}
\end{table}

The Karma design framework of parameters is the result of a systematic comparison of the Karma applications shown in Table~\ref{table_comparison}.
The design parameters are presented in Table~\ref{table_design_parameters}, and can be grouped into three categories: currency, interaction, and transaction.
As it turned out in many of the Karma applying works, a major design complexity is choosing the right amount of Karma currency in circulation~\citep{article_scrips,article_185,article_265,article_z02}. 
If there are too few currency units, there will be hoarding to save the scarce currency for very urgent situations to consume; 
if there are too many currency units, the value of a single value is not sufficient anymore to stimulate provision of resources.
In case of a time-variant resource supply, a dedicated amount control becomes necessary.
In the following, each of the design parameter groups is discussed in detail.


The currency parameters include parity, balance limits, amount control, initialization, and redistribution:
\begin{itemize}
    \item The parity represents the relationship between Karma and the resource. 
    The parity could be a price, meaning that resources can be traded for different amounts of Karma. 
    This can be useful in case of non-homogeneous resources, e.g. a large file can be exchanged against three small files.
    The parity could also be binary, meaning that one resource can be traded for exactly one unit of Karma. 
    This can be useful for homogeneous, atomic resources, e.g. one evening of babysitting service.
    The parity could also be a threshold, meaning that an agent needs a certain amount of Karma to be eligible to consume resources.
    
    \item The balance limits represent the limitations of Karma ownership. 
    Either, agents could have unlimited amounts of Karma, or there could be restrictions towards upper bounds to avoid hoarding. 
    Depending on the resource allocation mechanism, the absence of lower bounds could also enable an agent to have Karma debts to a certain amount.
     
    \item The amount control represents the monetary control mechanism to control the total amount of Karma circulating in the system. 
    Either it could be controlled by the system itself, in order to keep the amount of Karma per capita at specific levels.
    It could also just be constant or not controlled at all. 
    Besides, Karma points could also expire or depreciate. 
    
    \item The initialization represents how agents are initially provisioned with Karma. 
    This could mean at the beginning of the establishment of a Karma mechanism, or dynamically for new joining agents. 
    This could either happen by an equal, or random initial endowment, or not at all, meaning that agents need to provide some times before they can consume. 
    \cite{article_foodbank} even used a weighted initial endowment according to the central coordinators perception of need.

    \item The redistribution represents schemes how Karma is redistributed across agents. 
    It could be, that after each period, Karma is distributed via taxation of property.
    It could also be redistributed by taxation of payments, via a lottery, removal via expiry, or there could be no redistribution at all.
\end{itemize}

The interaction parameters include price control, price limits, resource provision, resource allocation, counter-party, peer selection, decision-making, urgency process, and temporal preference:
\begin{itemize}
    \item The price control represents how prices for resources are determined. 
    Prices could be determined by market mechanisms such as auctions, where forms of bidding processes take place with varying degrees of information transparency. Besides, prices could be determined by a central coordinating authority (the system), or not determined as all (in case of binary parity). 

    \item The price limits represent the limitations of resource prices. 
    Prices could be not limited at all, meaning also negative prices could be possible leading to earning Karma through consumption (e.g. goods with negative utilities). 
    Prices could also only be positive, defined by rational behaviour, or prices could be fixed (in case of binary parity).

    \item The resource provision represents who is the resource provider. 
    Either all agents could potentially provide resources, or only one agent could provide resources, or the network or system itself.

    \item The resource allocation represents how a decision is made whom to provide the resource to. 
    Depending on the price control, it could be highest or the second highest bidder, who receives the resource; it could also be a trading \& exchange system that executes orders such as it is the case in double actions. 
    In cases, where not a large number of agents, but only a pair of two agents interact, the resource provider could assess if the bid is high enough the effort. Besides, if the resource is provided by a central system or network, it could always be provided to the agent, in case it accepts the system defined price. 

    \item The counter-party represents with whom an agent interacts. 
    It could be that the agent interact with (all) other agents in case of a market (auction). 
    Besides, it could also be that the agent just interacts with exactly one other agent, or with no other agent but the system itself.

    \item The peer selection represents how agents find their peers for an interaction. 
    It could be that agents find (all) other peers through the market, so they don't decide actively for peers. 
    Similarly, the neighbourhood or a recommending/guidance system could provide a subset of the market to compete with. 
    It could also be, that agents are randomly assigned through each other, e.g. meeting at an intersection. 
    And finally, it could be actively selecting peers. 

    \item The decision-making represents if agents are free in making their decisions. 
    It is important to emphasize, that in the Karma mechanism, the buyers and sellers are always free to make their decisions (neglecting their urgency and needs). 

    \item The urgency process represents how the urgency of agents emerges over time. 
    It could be that all agents share a similar (homogeneous) or a different (heterogeneous) urgency process, with similar or different probabilities for different levels of need. 
    Moreover, it could be that the urgency at a later point in time depends on previous resource allocation, e.g. starvation or dehydration.
    
    \item The temporal preference represents how agents prefer present over future consumption. 
    It could be, that all agents share a similar temporal consumption preference (homogeneous), or that they differ in that (heterogeneous).  

\end{itemize}

The transaction parameters include payment amount, payment receiver, Karma gain, and Karma loose:
\begin{itemize}
    \item The payment amount represents what an buyer (agent) needs to pay in case it gets the resource allocated. 
    Depending on the price control and other design parameters, there are many payment rules possible. 
    The buyer could pay the bid, the bid of the peer, the difference between its and its peer's bid, a fixed price (in case of binary parity) or even nothing.

    \item The payment receiver represents who receives the Karma paid by the buying agent. 
    It could be, that the resource providing agent receives the payment, or that the payment is equally distributed across the population of agents, or weighted distributed according to how much Karma the population has, or to the system.

    \item The Karma gain represents how agents can earn their Karma units. 
    Agents could earn Karma by providing resources; in case of negative prices it could also happen by consumption.

    \item The Karma loose represents how agents can loose their Karma units. 
    Agents could loose Karma by resource consumption or expiry. 
    Besides, agents could also loose Karma by untrustworthy, rule-violating behaviour.
\end{itemize}

Karma was not only applied as a resource allocation mechanism, but also in a variety of many other applications and contexts.
Table~\ref{table_contexts} summarizes the application fields of Karma in the different literature fields.
In Section~\ref{secA1} of the appendix we summarize relevant terminologies to describe these applications, as well as we provide alternatives to Karma for each of these application fields.

\begin{table}[H]
    \caption{\textbf{Karma in different contexts}}
    \label{table_contexts}
    \begin{tabularx}{\textwidth}{ll}
        \toprule
        \textbf{Literature Field} & \textbf{Application Field} \\
        \midrule
        \textbf{Blockchain}  & \\
            \hspace*{0.5cm}           & Reputation / credibility system \\
            \hspace*{0.5cm}           & Trust system \\
            \hspace*{0.5cm}           & Secure accounting system \\
            \hspace*{0.5cm}           & Micropayment system \\
            \hspace*{0.5cm}           & Trading system \\
            \hspace*{0.5cm}           & Information dissemination technology \\
            \hspace*{0.5cm}           & Scrip system \\
            \hspace*{0.5cm}           & Token economy \\
            \hspace*{0.5cm}           & Credit scheme \\
            \hspace*{0.5cm}           & Crypto currency \\
            \hspace*{0.5cm}           & Lightweight currency \\
            \hspace*{0.5cm}           & Technological prerequisite for... \\
            \hspace*{0.5cm}           & \hspace*{0.5cm} smart contracts \\
            \hspace*{0.5cm}           & \hspace*{0.5cm} distributed hash tables \\
            \hspace*{0.5cm}           & \hspace*{0.5cm} distributed ledger \\
            \hspace*{0.5cm}           & \hspace*{0.5cm} minting proof-of-work \\
        \midrule
        \textbf{Network \& Technology}  & \\
            \hspace*{0.5cm}           & Network Protocol \\
            \hspace*{0.5cm}           & Self-Coordination in peer-to-peer computer networks \\
            \hspace*{0.5cm}           & Solution malicious behaviour in peer-to-peer networks... \\
            \hspace*{0.5cm}           & \hspace*{0.5cm} free-riding problem \\
            \hspace*{0.5cm}           & \hspace*{0.5cm} hidden actions problem \\
            \hspace*{0.5cm}           & \hspace*{0.5cm} lotus eater attack \\
            \hspace*{0.5cm}           & \hspace*{0.5cm} Sybil attack \\
            \hspace*{0.5cm}           & \hspace*{0.5cm} Eclipse attack \\
            \hspace*{0.5cm}           & \hspace*{0.5cm} spoofing attack \\
        \midrule
        \textbf{Filesharing}  & \\
            \hspace*{0.5cm}           & Computation Resource Sharing \\
            \hspace*{0.5cm}           & File Sharing \\
        \midrule
        \textbf{Economics}  & \\
            \hspace*{0.5cm}           & Self-contained economy \\
            \hspace*{0.5cm}           & Non-monetary market \\
        \midrule
        \textbf{Game Theory}  & \\
            \hspace*{0.5cm}           & Dynamic population game \\
        \midrule
        \textbf{Behaviour}  & \\
            \hspace*{0.5cm}           & Incentive mechanism \\
            \hspace*{0.5cm}           & Fairness enforcement \\
            \hspace*{0.5cm}           & Altruism enforcement \\
            \hspace*{0.5cm}           & Contribution enforcement \\
        \bottomrule
    \end{tabularx}
\end{table}

\section{Future Research Directions}\label{future}

We foresee four groups of promising future research questions: 
(i) novel applications, 
(ii) extensive in-depth analysis of the Karma mechanism, 
(iii) effect of Karma agent penetration on systems, and 
(iv) an economic, comparative analysis of Karma with monetary market mechanisms.

Karma is a resource allocation mechanism, that could be used in many further applications. While applications were already published in computation power resource networks, telecommunication networks, and road transportation networks only, we foresee numerous, possible future research endeavours that could revise, trading on renewable energy markets (smart grids), internet of things applications, internet traffic management, logistics (container terminal and warehouse management), production and supply chains within corporate organizations, and ad-hoc, emergency networks. 
Besides, the field of transportation networks and telecommunication networks provide various research topics, such as different transportation modalities including resource allocation problems on trucks, on rails, on ship, and in the air; or different problems that internet service providers face.

In terms of the extensive, in-depth analysis of the Karma mechanism, one could analyse into more detail 
(i) the robustness to population heterogeneity, 
(ii) the fairness and efficiency, 
(iii) the on-line learning and application of (sub-)optimal policies by agents, 
and (iv) the out-of-equilibrium behaviour.
Besides the theoretic computations of the optimal, selfish strategy of agents, it would be interesting to assess whether (human) agents are actually able to learn the optimal strategy themselves and to act accordingly. 
Moreover, how would non-optimal behaviour (policies) affect the efficiency and fairness aspect that Karma proposes as value?
In terms of the aspect of fairness, it would be interesting to analyse the effect of the Karma mechanism on the fairness in depth or at certain aggregation levels. 
In addition to that, extensive works on the interplay and relationship between fairness and efficiency are promising.
As \cite{article_600} argue, first analysis on the robustness of the Karma mechanism to population heterogeneity was conducted, however, more work needs to be done in this aspect of the Karma mechanism as well.

For many real world applications it is reasonable to assume, that during the introduction of Karma based systems, there will be transition periods where populations of Karma-using and not-Karma-using agents will coexist.
The effects of Karma on fairness and efficiency for mixed-population systems at different penetration rates could be interesting to study.

In addition to that, it would be interesting to systematically compare and analyse Karma as a non-monetary mechanism with monetary mechanisms. 
Karma could be used to comparatively measure the efficiency losses of translating utilities to money in monetary markets.
Besides, it could be interesting to quantify boundaries for market conditions in which non-monetary Karma markets outperform monetary markets.

\section{Methodology}\label{methods}


\subsection{Literature Review with Systematic Mapping Study}

Literature reviews are secondary studies (meta studies), that represent a method to condense, synthesize, and to survey knowledge from a large collection of primary studies.
Conducting a literature review can serve a multitude of reasons: to present the current state of knowledge on a topic, to identify all evidence for or against a research question, to establish where there is consensus or debate on an scientific issue, to show how knowledge and models around a topic evolved historically, and to look out for future research directions~\citep{litreview_A}.

\begin{figure}[H]
    \centering
    \includegraphics[width=10.5 cm]{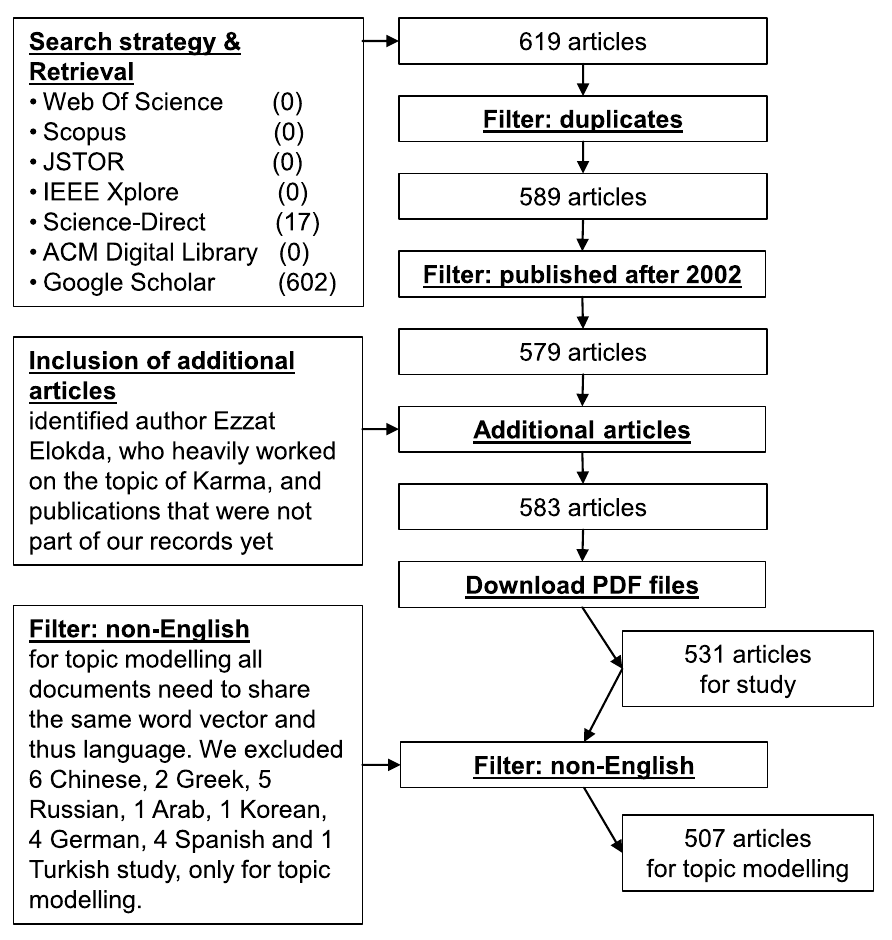}
    \caption{
        \textbf{Article search \& selection process.} 
        }
    \label{fig_SMS}
\end{figure}

Four cognitive biases can affect the scientific integrity of literature review and lead to systematic errors~\citep{litreview_D}.
The author bias (confirmation bias) describes the threat to not include all relevant studies, and to include studies in favor of the expected outcome only.
The publication bias is the risk of laying too much emphasis on studies with positive results (which are usually more likely to get published) and neglecting studies with negative or inconclusive results (that often end up unpublished or not published in journals).
The English language bias refers to the neglect of non-English literature.
The non-reproducibility bias is the risk of not following a formal, systematic approach, which impedes independent reproduction of conclusions by others.

The three types of literature review differ in their approach, purpose, and scope~\citep{litreview_B}.
While the \textit{traditional literature review} relies on the authors' knowledge, experience \& network, 
the \textit{systematic literature review} requires a formalized, systematic, and scientific approach to discuss a topic in a well-balanced manner.
The \textit{systematic mapping study} follows a formalized process as well, but allows for a broader scope with more general questions when reviewing the topic of interest. This broader review of a larger literature corpus can include quantity, type, origin, and temporal scope of primary studies as well, and is usually presented in a map.

\subsection{Research Protocol}

The systematic mapping study for this literature review was conducted in the five step approached based on \cite{litreview_K}.
Figure~\ref{fig_SMS} explains the article search and study selection procedure quantitatively.

\subsubsection{Protocol Planning}

\textbf{Motivation: }
Starting from the original paper~\citep{vishnumurthy2003karma}, Karma recognized attention in many different fields and applications. 
With regard to the number of primary and secondary citations it received and still receives, the interest in Karma is growing. 
New applications and significant adoptions can be found every year. 
However, the citing literature is dispersed over time and across disciplines:
While at the beginning Karma was a sore subject in the filesharing literature (to overcome free-riding in peer-to-peer networks), 
it has become more prominent in the block-chain literature (foundational technology for modern crypto currencies), 
and is about to receive significant attention in the game theory and control literature (as fair and efficient resource allocation mechanism). 
There is a lack of an interdisciplinary, comprehensive survey on Karma and all its applications yet, that lays the foundation for systematic, future research.
Therefore, this study aims to provide a comprehensive, interdisciplinary, systematic mapping study on Karma as a concept in different fields, and especially as a resource allocation mechanism.

\textbf{Scope:}
We are interested in understanding the structure of the literature that evolved around the first work on Karma in the filesharing literature~\cite{vishnumurthy2003karma}, in order to better understand how Karma is defined in different fields, and to establish a common, cross-domain language. 
Moreover, we focus on Karma as a resource allocation mechanism, and systematically analyse differences and similarities of applications. 
In addition to that, we thematize how the game theoretic formalism can be used to quantitatively analyse the Karma mechanism.
Finally, we aim to elaborate on potential future research for Karma. The scope of this study is summarized by following ten research questions:
\begin{enumerate}
    \item How did Karma evolve in the literature over time?
    \item What is the intellectual field-structure of the Karma citing literature corpus?
    \item What is Karma defined as and which relevant terminology exists in these fields?   
    \item Which concepts as an alternative to Karma exists in these fields?
    
    \item How is Karma different from other resource allocation mechanisms?
    \item How is Karma applied as resource allocation mechanism?
	\item What are the parameters \& options  when designing Karma applications?
	\item How to model and formally describe the Karma mechanism?
	\item What is a rational policy in a Karma game?
	\item Which features does the Karma mechanism provide?
    \item What are promising future research questions for Karma?
\end{enumerate}

\textbf{Search strategy:}
We search for all papers, that cite the original paper on Karma~\citep{vishnumurthy2003karma}. 
For this purpose, we review following established, frequently used, international, interdisciplinary, scientific literature databases: Web Of Science, Scopus, JSTOR, IEEE Xplore, Science-Direct, ACM Digital Library, and Google Scholar.
The results are manually reviewed to account for duplicate records.
\endnote{The database queries were conducted on October 1st, 2023.}

\textbf{Selection criteria:}
We include any record that cites the original paper in its reference list.
We exclude any record from the citing literature that was published before 2002, as the original paper was published 2003 itself, and might have been cited as a forthcoming working paper version one year before that.

\subsubsection{Literature Retrieval:}
The original paper was published as a workshop paper at Cornell university \textit{Vishnumurthy, V., Chandrakumar, S., \& Sirer, E. G. (2003, June). Karma: A secure economic framework for peer-to-peer resource sharing. In Workshop on Economics of Peer-to-peer Systems (Vol. 35, No. 6)}. 
As it was not published at a conference or journal later on, solely Science-Direct (17 hits) and Google Scholar (602 hits) returned records. 

\subsubsection{Study Selection \& Quality Assessment}
For each of the records, we captured title of the publication, link to the publications website, the number of citations (according to Google Scholar), and download-link to a PDF (available for 441 records). 
We then manually complemented the records with missing download-links.
After manual inspection of the records, we removed duplicates and remain with 589 unique papers citing the original work. 
These duplicates occurred due to multiple versions of the same paper (e.g., preprints) indexed in Google Scholar. 
For each of the records, we (i) download the scientific publication as a PDF document from the download-link, (ii) download their bibliography meta-data as Bibtex file from Google Scholar, (iii) extract the full-text from the PDF files into TXT files for further processing, and (iv) determined how often Karma was mentioned in the papers by counting the term "Karma".
This procedure was possible for 531 papers of the citing literature, the others were not available for download and could not be found anywhere else. 
We used the file size of PDF files and number of characters in the TXT files for manual inspection to identify downloaded files that were empty or just abstracts (too short for a publication). 
Several of the download links from Google Scholar were not working and had to be complemented manually. 
The procedure was repeated iteratively until we remained with all data for the 531 records.
In terms of document quality assessment, we choose to include all studies.
Similar to the original Karma paper, many citing works were published outside of journals and conferences, for example on arxiv or as university-intern publications; therefore, we are convinced there are no issues with the inclusion of all the records we found. 

\subsubsection{Classification \& Analysis}


The classification \& quantitative analysis was conducted by topic modelling using the Latent Dirichlet Allocation algorithm.
During text-preprocessing we removed all special, not alphanumeric characters, transformed words to lower case, removed easy and stop words, and applied Porter's stemming method~\cite{antons2019content,porter1980algorithm}. 
In a manual, iterative process, we reviewed the most frequent 1-grams, 2-grams and 3-grams, and where appropriate, excluded from the text or combined the n-grams to single words. 
In total we found 55 1-grams, 435 2-grams and 51 3-grams. 
The processing of n-grams was instrumental in improving our models and creating meaningful topics, as consecutive terms like "game" and "theory" should be considered as one word.
We excluded 24 documents from topic modelling, as they were not in English language and therefore did not share the same word vectors (6 Chinese, 5 Russian, 4 German, 4 Spanish, 2 Greek, 1 Arab, 1 Korean, 1 Turkish).

We then used a software implementation~\endnote{Python package tomotopy (v.0.9.0, \url{https://bab2min.github.io/tomotopy/v0.5.1/en/}).} for calculating the LDA models. 
We rendered around 1338 models with different hyper parameter combinations. Consistently, for different alpha and beta values, an inherent optimal number of topics of around 30 was identified. After manual inspection of the topics and topic-word-distribution, we aggregated the topics resulting in six distinct and meaningful topics using grounded theorizing: "Behaviour", "Game Theory", "Economics", "Filesharing", "Network \& Technology", and "Blockchain". 
Our finally selected model was trained based on 503 documents with in total 5,264,546 words, covering 86,794 different words of vocabulary (entropy of words -7.81), using an initial alpha of 0.07 (after training the alpha of each topic was around 0.35) and beta of 0.01. 
The final model has a perplexity of 4048.81, and a negative likelihood of -8.31.

\subsubsection{Mapping}

First, we analyse the number of primary and secondary citations over time and how often the term "Karma" occurs in each paper of the citing literature corpus. 
Then, we use the topic models to analyse, in which fields Karma was applied most over time.
Finally, we employ the topic models to analyse the similarity of topics and render word clouds to understand the most frequently used words of each topic.

\subsection{Topic Modelling with Latent Dirichlet Allocation}

Topic modelling is a mathematical approach to study large datasets, that can be understood as a principal component analysis, that enables the user to find meaningful topics and to uncover latent structures in large data sets. 
While topic modelling was successfully applied in a variety of data including images, population genetics, survey data and social networks, its most prominent application can be found in bibliometric studies, when analysing textual data~\citep{hannigan2019topic}.
One of the most recognized algorithms for topic modelling is Latent Dirichlet Allocation (LDA)~\citep{blei2003latent}. 
It assumes that, for a given data set, there is an inherent, latent, intellectual structure which follows a Dirichlet distribution. 
In applications with text, LDA inputs a textual corpus matrix, that consists of the word vectors of each document, 
and outputs two probabilistic matrices, that represent a topic-document-distribution, and a topic-word-distribution. 
The quality of topic models can be assessed by the two metrics perplexity and negative likelihood~\citep{blei2003latent}.
Prerequisites to LDA are text-preprocessing and hyper parameter selection.

During text-preprocessing, a text is split into single words, a list of unique single words is determined, and the word vectors represent the number of times the unique words occur in a document. 
For text-preprocessing, it is common to transform all words to lowercase, to remove easy and stop words, to apply Porter's stemming method, and to identify and replace n-grams~\cite{antons2019content}.

Hyperparameters, such as alpha (how concentrated is the topic-document-distribution) and beta (how concentrated is the topic-word-distribution) of the Dirichlet distribution, and the number of topics expected to find, must be preset upfront. 
One common suggestion followed in the literature for the choice of alpha and beta is to choose alpha as ratio of 50 and the number of topics, and beta as 0.1. 
The number of topics can either be predefined or considered as inherent to the corpus. 
In the latter case, it is common to minimize negative likelihood respectively maximize perplexity of the models in order to find the optimal number of topics~\citep{griffiths2004finding}.

\section{Concluding Remarks}\label{conclusion}

Karma is a non-monetary, artificial currency for resource allocation;
it can only be gained by producing and only be lost by consuming a specific resource.
Karma is Pareto-efficient, fairness-enhancing, near incentive-compatible, robust towards population heterogeneity, and can also lead to a decrease in resource scarcity in peer-to-peer markets.

This study set out to provide an exhaustive literature review on the Karma mechanism for resource allocation, and to connect Karma with the context of its literature.
A systematic mapping study and topic modelling was conducted, finding that Karma originated as a concept in the filesharing literature, that gained momentum in six different disciplines.
Karma was applied as a resource allocation mechanism in various applications, which have been systematically compared.
As a result, this study provided an instrumental framework with design parameters and options, to enable more systematic application of the Karma mechanism.
Finally, the results from the review were used to recommend future research directions for Karma.



\vspace{6pt} 




\authorcontributions{
    \textbf{Kevin Riehl: } Conceptualization, Methodology, Software, Formal analysis, Investigation, Data Curation, Writing - Original Draft, Visualization, Project administration. 
    \textbf{Anastasios Kouvelas: } Writing - Review \& Editing, Validation.
    \textbf{Michail Makridis: } Writing - Review \& Editing, Validation, Supervision.
}

\funding{
    This research received no external funding.
}

\dataavailability{
    The literature corpus (list of publications considered for this review) and computational results of the Latent Dirichlet Allocation topic modelling can be found in the online GitHub repository: \href{https://github.com/DerKevinRiehl/karma_literature_review}{https://github.com/DerKevinRiehl/karma\_literature\_review}.
} 




\acknowledgments{
    We would like to thank Ezzat Elokda, Florian Dörfler, and Saverio Bolognani for the valuable and useful feedback and comments, which was instrumental when improving this work.
}

\conflictsofinterest{
    The authors declare no conflicts of interest.
} 





\newpage
\appendixtitles{yes} 
\appendixstart
\appendix

\section{Tables}\label{secA0}

\begin{table}[ht!]
    \caption{\textbf{Applications of Karma as a network resource allocation mechanism (1 / 2) }}
    \label{table_applications_A}
    \includegraphics[width=\linewidth]{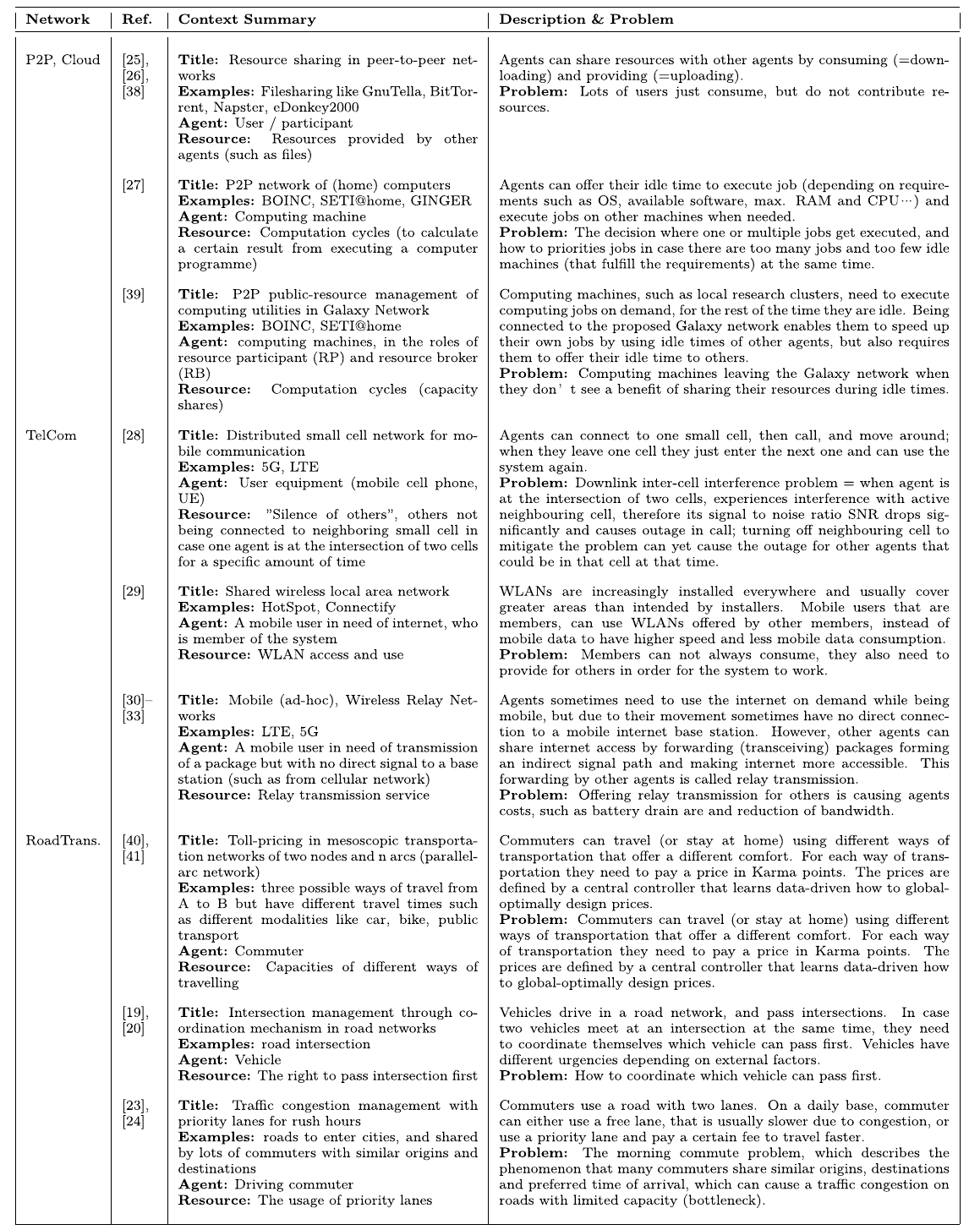}
    \begin{tabular}{p{\linewidth}}
    \; \\
    \end{tabular}
\end{table}

\begin{table}[ht!]
    \caption{\textbf{Applications of Karma as a network resource allocation mechanism (2 / 2) }}
    \label{table_applications_B}
    \includegraphics[width=\linewidth]{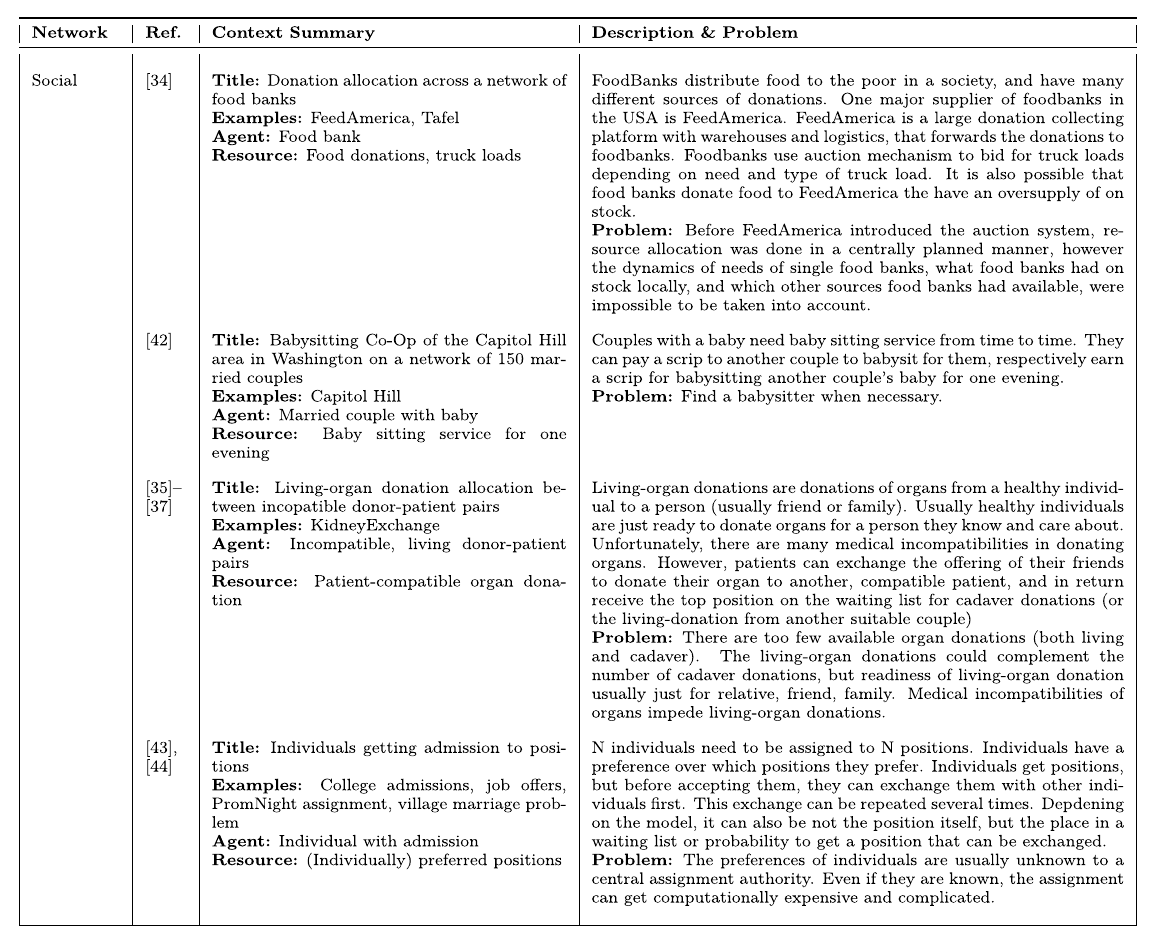}
    \begin{tabular}{p{\linewidth}}
    \; \\
    \end{tabular}
\end{table}

\begin{table}[ht!]
    \caption{\textbf{Resource source, goals and actions in Karma applications}.
    }
    \label{table_applications_C}
    \includegraphics[width=\linewidth]{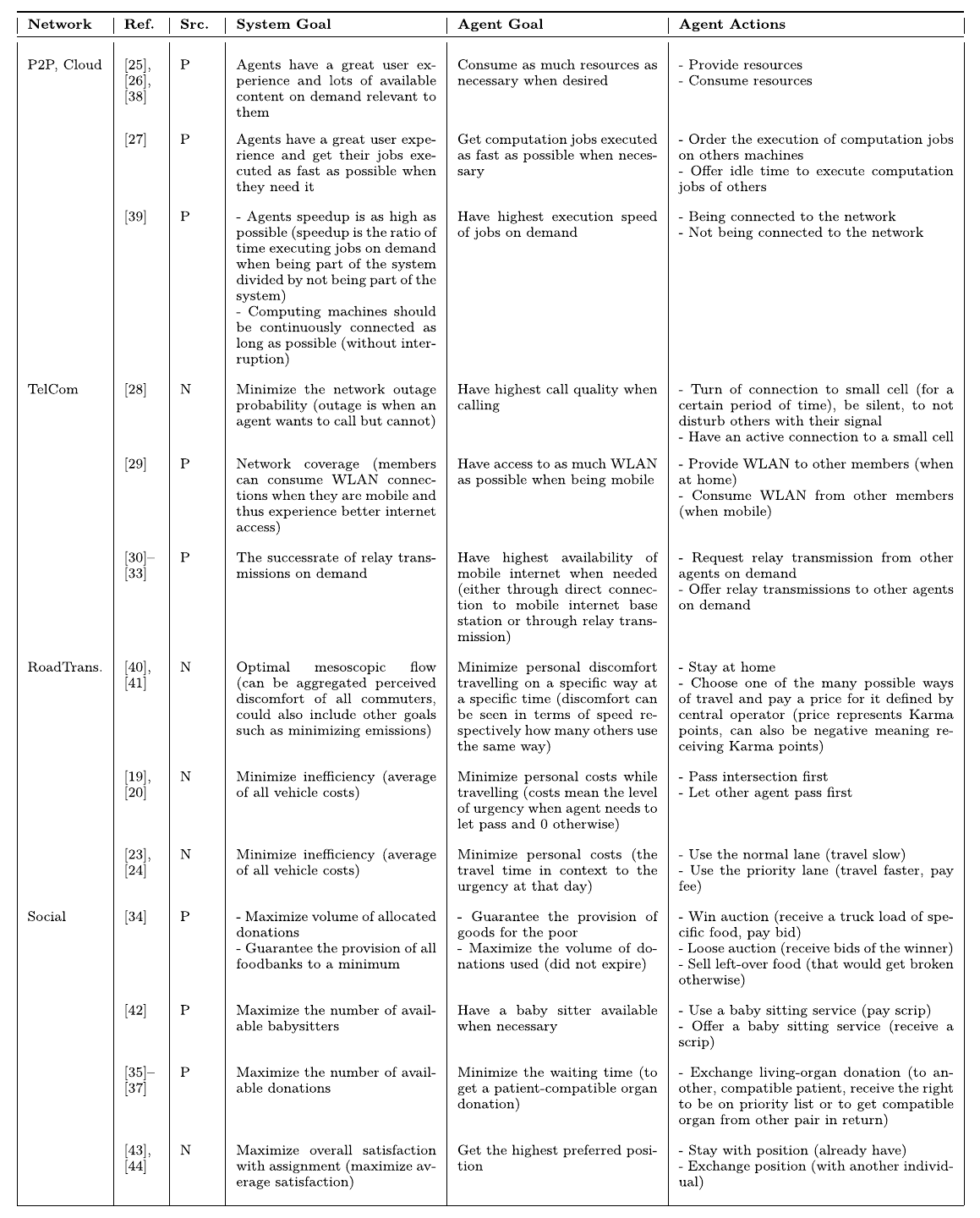}
    \begin{tabular}{p{\linewidth}}
    \; \\
    \end{tabular}
    \noindent{\footnotesize{* Column "Src." describes the source of resources, where P stands for resource generation by participants and N stands for resource inherence by the network.}}
\end{table}

\begin{table}[ht!]
    \caption{\textbf{Comparison of Karma applications}.
    }
    \label{table_comparison}
    \includegraphics[width=\linewidth]{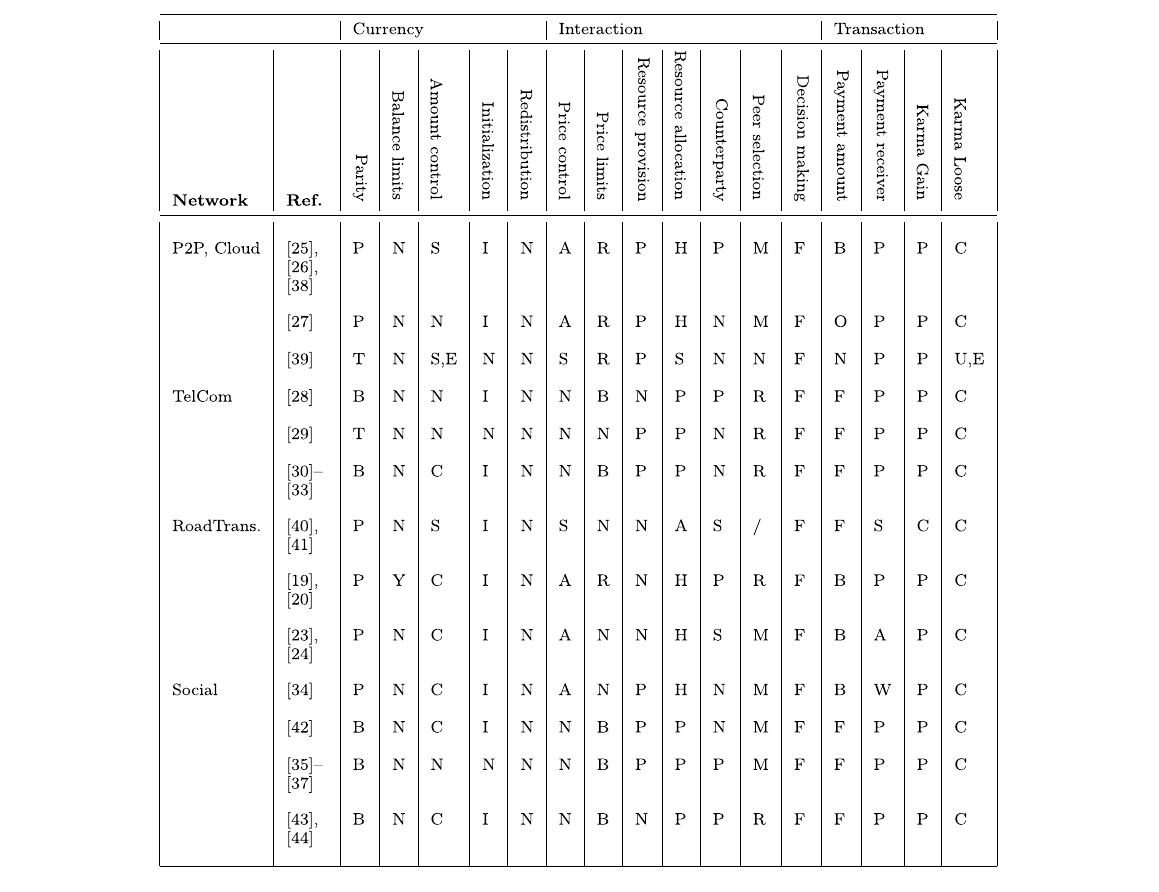}
    \begin{tabular}{p{\linewidth}}
    \; \\
    \end{tabular}
    \noindent{\footnotesize{* This table shows the different Karma design parameters for the applications. The following explains the abbreviations in this table of different options for the design parameters. 
	The \textit{parity}: P = price (resources can be traded for different amounts of Karma); B = binary (one resource can be traded for one unit of Karma); T = threshold (one needs a certain amount of Karma to be eligible to consume resources). 
	The \textit{balance limits}: N = no (agents can have as many or as few Karma points as they want); Y = yes (agents can have an amount of Karma which is limited by an lower bound and a defined upper bound). 
	The \textit{amount control}: S = control by the system over time; N = no control; E = expiration of Karma units; C = constant number of Karma units. 
	The \textit{initialization}: I = equal initial endowment; N = no initial endowment. 
	The \textit{redistribution}: N = none. 
	The \textit{price control}: A = a market-like price determination, auctions; S = system controlled prices; N = no price determination (in case of binary parity).	  
	The \textit{price limits}: R = rational (prices above zero); B = binary (only one unit of Karma); N = none (all prices, including negative prices). 
	The \textit{resource provision}: P = an agent; N = the network / system. 
	The \textit{resource allocation}: S = system decides; P = up to the provider; A = always guaranteed to anyone who pays. 	
	The \textit{counterparty}: P = pairs (with one other agent); N = multiple or all other agents (e.g. market, auction); S = the central system coordinator (e.g. system defined prices). 
	The \textit{peer selection}: M = market; N = a local neighbourhood / or a submarket suggested respectively guided by the system; R = random encounters that cannot be controlled by consumer and provider. 
	The \textit{decision-making}: F = agents are free to make their decisions (neglecting their urgency and needs). 
	The \textit{payment amount}: B = bid (used in auctions); O = order (system will calculate the costs); N = nothing; F = fix (fixed price or binary parity). 
	The \textit{payment receiver}: P = resource provider (resource provider receives); S = system (system receives); A = equal redistribution (to all participants); W = weighted redistribution (to all participants). 
	The \textit{Karma gain}: P = provision (earn Karma by providing resources); C = consumption (earn Karma by consuming in case of negative prices). 
	The \textit{Karma loose}: C = consumption (loose Karma by consuming resources); E = expiration of Karma points; U = untrustworthy behaviour (system rule violation).}}
\end{table}

\clearpage

\section{Karma \& Alternatives}\label{secA1}

(see attached online appendix PDF file)

\begin{adjustwidth}{-\extralength}{0cm}
\printendnotes[custom] 

\reftitle{References}


\bibliography{bibliography}

%


\PublishersNote{}
\end{adjustwidth}
\end{document}